# Experimental observation of one-dimensional motion of interstitial skyrmion in FeGe

Dongsheng Song[1,2], Weiwei Wang[1,2], Jie-Xiang Yu[3], Peng Zhang[4], Sergey S. Pershoguba[5], Gen Yin[6], Wensen Wei[2], Jialiang Jiang[2,7], Binghui Ge[1], Xiaolong Fan[4], Mingliang Tian[2,7,8], Achim Rosch[9], Jiadong Zang[5,9,10*], and Haifeng Du[1,2,7*]

[1]Institutes of Physical Science and Information Technology, Anhui University, Hefei 230601, China

[2]Anhui Key Laboratory of Condensed Matter Physics at Extreme Conditions, High Magnetic Field Laboratory, HFIPS, Anhui, Chinese Academy of Sciences, Hefei 230031, China.

[3]School of Physical Science and Technology, Soochow University, Suzhou 215006, China

[4]Key Laboratory for Magnetism and Magnetic Materials of the Ministry of Education, Lanzhou University, Lanzhou 730000, China

[5]Department of Physics and Astronomy, University of New Hampshire, Durham, New Hampshire 03824, USA

[6]Department of Physics, Georgetown University, Washington, D.C. 20057, USA

[7]Science Island Branch of Graduate School, University of Science and Technology of China, Hefei, Anhui 230026, China

[8]School of Physics and Materials Science, Anhui University, Hefei, 230601, China

[9]Institute for Theoretical Physics, University of Cologne, 50937 Cologne, Germany

[10]Materials Science Program, University of New Hampshire, Durham, New Hampshire 03824, USA

*Corresponding author: Jiadong.Zang@unh.edu and duhf@hmfl.ac.cn

**The interplay between dimensionality and topology manifests in magnetism via both exotic texture morphology and novel dynamics. A free magnetic skyrmion exhibits the skyrmion Hall effect under electric currents. Once it is confined in one-dimensional (1D) channels, the skyrmion Hall effect would be suppressed, and the current-driven skyrmion speed should be boosted by the non-adiabatic spin transfer torque $\beta$. Here, we experimentally demonstrate that stripes of a spatially modulated spin helix serve as natural 1D channels to restrict skyrmion. Using FeGe as a benchmark, an interstitial skyrmion is created by geometry notch and further moves steadily without the skyrmion Hall effect. The slope of the current-velocity curve for 1D skyrmion is enhanced almost by an order of magnitude owing to a large $\beta$ in FeGe. This feature is also observed in other topological defects. Utilizing the 1D skyrmion dynamics would be a highly promising route to implement topological spintronic devices.**

Chiral magnets are magnetic materials with broken inversion symmetry[1]. Enabled by the spin-orbit coupling, the emergence of the Dzyaloshinskii-Moriya interaction leads to exotic spin textures such as the magnetic skyrmion and spin helices[2,3]. Magnetic skyrmions are particle-like spin textures with non-trivial topology characterized by the integer valued topological charge $Q = \frac{1}{4\pi}\int \boldsymbol{n} \cdot (\partial_x \boldsymbol{n} \times \partial_y \boldsymbol{n})\, dxdy$, where $\boldsymbol{n}(x,y)$ is the unit vector along the spin direction. Nonzero $Q$ is critical to novel current driven dynamics of the skyrmion driven by the spin-transfer-torques (STTs)[4–6]. In a homogenous background, such as ferromagnetic or conical phases, motions on the skyrmion plane are collective low-lying Goldstone modes. Under an electric current, such collective dynamics can be concisely described by Thiele's equation[4,5,7] (**Supplementary Note I**),

$$\mathcal{D}(\beta\mathbf{u} - \alpha\mathbf{v}) + \mathbf{G} \times (\mathbf{u} - \mathbf{v}) + \mathbf{F} = 0 \quad (1)$$

Here $\mathbf{v}$ is the skyrmion velocity, and the effective electron velocity $\mathbf{u}$ relates to the current density $\mathbf{j}$ by $\mathbf{u} = -\frac{gP\mu_B}{2eM_s}\mathbf{j} = b\mathbf{j}$, where $g$ is the Landé factor, $\mu_B$ is the Bohr magneton, $M_s$ is the saturation magnetization, and $P$ is the polarization ratio[6]. The first term describes the dissipative force with the Gilbert damping $\alpha$, non-adiabatic STT $\beta$, and the skyrmion structure factor $\mathcal{D} = D\delta_{ij}$. The second term is the topological gyrotropic term with $\mathbf{G} = 4\pi Q\hat{z}$. Since both $\alpha$ and $\beta$ are small, the gyrotropic term is dominant, which leads to the celebrated skyrmion Hall effect[7–9] (**Fig. 1a**), and, furthermore, sets up the skyrmion velocity on the scale of $u$. Using high current density is seemingly the only way to achieve higher skyrmion speed, but it inevitably causes unwanted Joule heating of the sample and instability of the skyrmions.

Once the skyrmion is confined in a 1D channel, a paradigm shift takes place. The suppression of the skyrmion Hall effect indicates the diminish of the gyrotropic term. The remaining first term in Eq. (1) suggests a new speed scale $(\beta/\alpha)u$ would prevail[4,5] (**Supplementary Note I**). However, experimentally achieving such confined motion has been extremely difficult[10–20]. For example, even in nanoribbons, the transverse velocity of the skyrmion cannot be entirely suppressed. Even if succeeded, the defects and disorders at the sample boundary also significantly pin the skyrmion motion.

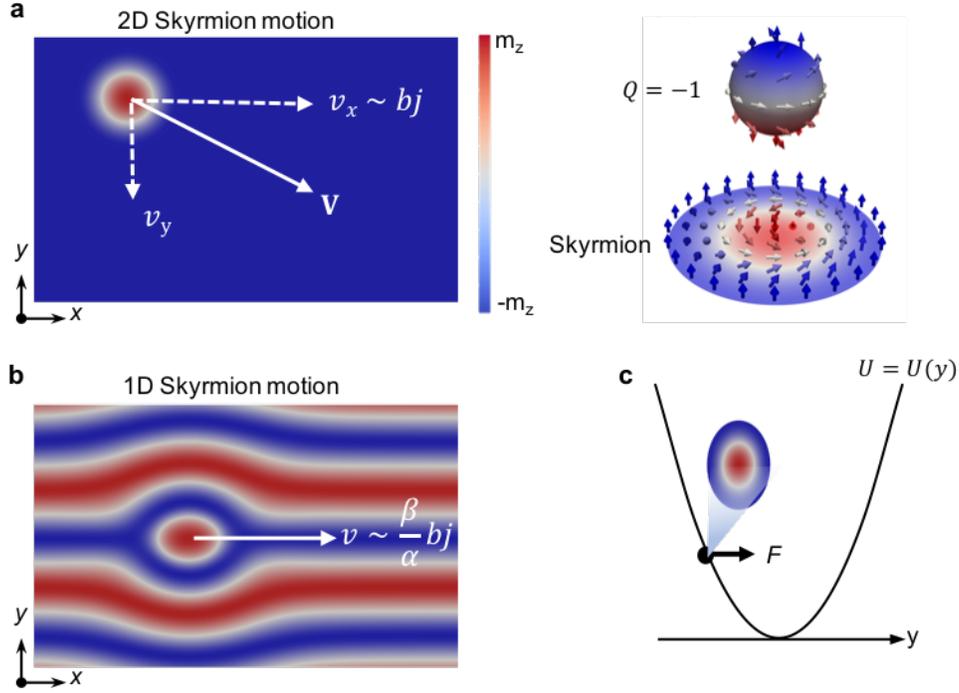

**Fig. 1 | Current-induced skyrmion dynamics. a,** Schematic of 2D skyrmion motion in the presence of electrical current. The white arrow marks the skyrmion trajectory, and the transverse motions represent the skyrmion Hall effect. The color bar represents the out-of-plane components of the magnetization. The spin configuration and topology of skyrmion are indicated on the left panel. **b,** An interstitial skyrmion moves without the skyrmion Hall effect owing to the extra potential raising from the helical background. The relationship between current density $j$ and skyrmion velocity $v$ is displayed for 1D and 2D cases, respectively, where $b = -\frac{gP\mu_B}{2eM_s}$ (see text for details). **c,** Schematic plotting of the quadratic potential of $U(y)$ for the interstitial skyrmion, which results in a force $F$ that balances the Magnus force when the interstitial skyrmion deviates from the equilibrium center.

A spin helix is a 1D-modulated magnetic configuration that serves as the ground state of chiral magnets[3]. The unique helical order has already demonstrated plenty of intriguing physical phenomena especially when coupled with conducting electrons, *e.g.* the emergent electromagnetic induction[21]. Owing to the 1D property, spin helix provides natural highways for skyrmion motion, while surface defects typically associated with nanostructures are nicely avoided[22]. By placing an individual skyrmion in the interstitial space of a spin helix (**Fig. 1b**), the combined texture survives at zero magnetic field. Although the translational symmetry is broken along the helical **q**-vector, it is still respected in the perpendicular direction. A straight motion of the skyrmion along the helical stripe is thus theoretically expected[22–24].

In this work, we experimentally demonstrate that a spin helix is an ideal platform for the 1D motion of skyrmion in a typical B20 chiral magnet FeGe. Under the action of electric current, the interstitial skyrmion is created at the specifically designed notch and subjected to an effective potential barrier from the surrounding distorted spin helix (**Fig. 1c** and **Supplementary Note II**). The skyrmion indeed moves right at the center between neighboring helix lanes. Particularly, the velocity of interstitial skyrmion shows a distinct relationship with respect to the current density, compared to that in the 2D homogeneous background.

**Fig. 2a** shows the schematics of a FeGe nanostripe fabricated by an in-house procedure on a self-customized electrical chip for in-situ transmission electron microscopy (TEM) experiments.[20,25] (**Methods** and **Extended Data Fig. 1**). The nanostripe's width, length, and thickness are 1 μm, 6 μm, and 150 nm, respectively. The two ends of the nanostripe were fixed by the Pt electrodes that were connected to a voltage source. Following the idea of creating skyrmions using geometrical defects in the ferromagnetic (FM) background[4], we carved a notch on one edge of the nanostripe. The electrical current pulse with a duration of 20 ns and a frequency of 1 Hz was applied along the long axis of the nanostripe. The in-situ Lorentz TEM was used to track the magnetic dynamics.

Under routine zero-field cooling from room temperature to 95 K, the nanostripe adopts a complex helical ordering mixing multi-**q** wavevectors (**Extended Data Fig. 2**). To develop a well-defined single-**q** helical background, we utilized the electrical current pulse to reorientate the helical stripes based on the edge-induced ordering mechanism[26]. After applying a current pulse with $j \sim 10.7 \text{ MA} \cdot \text{cm}^{-2}$, multi-**q** helical domains are destroyed, and various topological defects with finite topological charges are introduced, including disclinations, dislocations, and meron pairs[22,27] (**Extended Data Fig. 2**). Subsequent current pulses push out these topological defects from the right end of the nanostripe. Simultaneously, a clean spiral phase with **q** ⊥ **j** enters from the left side of the nanostripe. After a series of current pulses, a well-ordered single-**q** helical state is eventually formed (**Extended Data Fig. 2** and **Supplementary Video 1**).

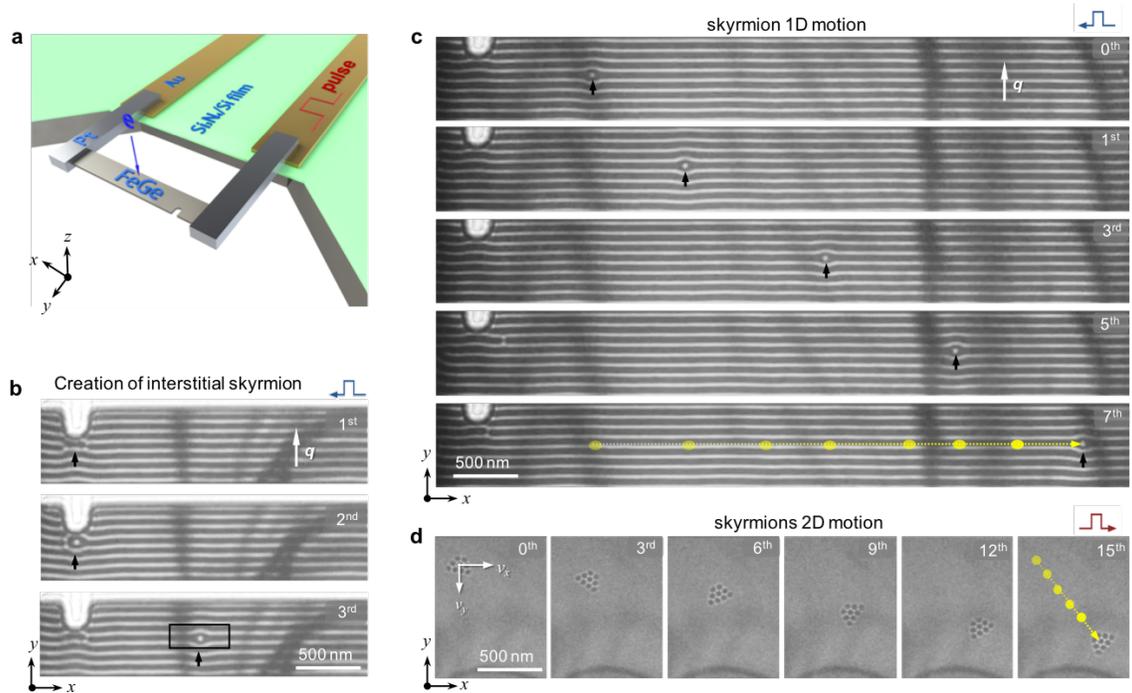

**Fig. 2 | Dimension-dependent skyrmion motion. a,** Schematic illustration of the FeGe microdevice. The left and right pads represent the Pt electrodes connected to the source of

nano-second current pulse through the Au wires in the Si$_3$N$_4$/Si film. A notch is carved at the upper left edge of the nanostripe. The nanostripe's width, length, and thickness are 1 µm, 6 µm, and 150 nm, respectively. The electron beam is along z direction. **b,** Representative snapshots of the nucleation of the interstitial skyrmion under the current pulse. The wavevector **q** of spin helix is indicated by the white arrow. The current density is ~ 11 MA·cm$^{-2}$ and the external magnetic field is $B = 0$ mT. The black arrows indicate the positions of interstitial skyrmion in spin helix. **c,** Representative snapshots of the motion of the interstitial skyrmion. The current density is ~ 9.7 MA·cm$^{-2}$. The black arrows indicate the positions of the interstitial skyrmion. The summary of the skyrmion trajectories under successive current pulses is schematically indicated with the yellow dots. **d,** The representative snapshots of the motion of skyrmion cluster in the ferromagnetic background under the external magnetic field of 120 mT. The current density is ~ 5.9 MA·cm$^{-2}$. Trajectory of the skyrmion cluster is summarized by the yellow dots.

Once established, the clean spin helix remains stable under the application of successive current pulses. In contrast, on account of the modified magnetic textures and nonuniform current distribution around the notch of the nanostripe, an individual skyrmion can nucleate right next to the notch (**Fig. 2b**). A subsequent persistent current pulse pushes that skyrmion in between well-defined helical lanes, and an interstitial skyrmion is formed[22] at a current density of $j \sim 11$ MA·cm$^{-2}$ (**Fig. 2b** and **Supplementary Video 2**). The local spin texture, analyzed by transport intensity of equation (TIE)[3] (**Extended Data Fig. 3**), shows that the interstitial skyrmion is elliptically flattened along the long axis, leading to the strong distortion of surrounded helical lanes. The unity topological charge of $Q = 1$ of the interstitial skyrmion is identified from the Lorentz Fresnel contrast. The electrical current could also excite other topological defects (*i.e.* the disclinations and meron pairs) as a random event at the notch under a relatively high current density of $j \sim$ 11 MA·cm$^{-2}$. Once one of them is created, it is very stable and could be effectively driven at lower current density without breakdown or introducing new defects.

Following the successful creation of the interstitial skyrmion, we explore its motion along the lanes of spin helix under varied current density. **Fig. 2c** shows the representative snapshots of the interstitial skyrmion motion at a typical current density of $j \sim 9.7$ MA·cm$^{-2}$. Clearly, the skyrmion moves only along the channel between two fixed helical lanes without the detrimental skyrmion Hall effect[7–9]. The background spin helix remains intact without breakdown in the whole process (**Supplementary Video 3**). The yellow spots summarize the trajectory of skyrmion motion under each current pulse in **Fig. 2c**. The deviation from equal spacing can be attributed to the random pining sites in the bulk of the sample[6,19,20,25]. The overall average displacement in **Fig. 2c** shows that the mean velocity $v$ of the interstitial skyrmion is up to $\sim 22$ m·s$^{-1}$. Upon reversing the direction of the electrical current, the interstitial skyrmion experiences the opposite driving force and moves in the opposite direction (**Extended Data Fig. 4** and **Supplementary Video 4**).

Such 1D motion persists in a reasonable range of driving current density. The lower critical current density of driving the interstitial skyrmion is about $j_{c1} \sim 7.1$ MA·cm$^{-2}$, below which defects pin the skyrmion. The upper critical current density is about $j_{c2} \sim 11.1$ MA·cm$^{-2}$, above which the ordered spin helices are destroyed (**Supplementary Video 5**) and the skyrmion disappears due to the Joule heating effect[20] and/or the too strong STT-induced skyrmion Hall effect[7–9]. In between these two critical currents, skyrmion flow-regime[4] is observed, and the interstitial skyrmion could be reproducibly driven back and forth with the vanishing skyrmion Hall effect. Occasionally, under a current close to $j_{c2}$, the interstitial skyrmion appears elongated along the current direction (**Extended Data Fig. 5** and **Supplementary Video 6**) or transform into a thermodynamically stable meron pair[22] (**Extended Data Fig. 6** and **Supplementary Video 7**). In the latter case, the interstitial skyrmion jumps into the lower nearest helical lane and forms a head-to-head meron. Such behavior can be attributed to the skyrmion Hall effect, where the increased Magnon force leads to the collapse of the interstitial skyrmion and thus a structure transformation occurs while the finite topological charge is still conserved[22].

As a comparison, the typical motion of skyrmions in the homogenous ferromagnetic background is presented (**Fig. 2d** and **Supplementary Video 8**). It is an apparent 2D motion with a clearly tractable skyrmion Hall angle of $\theta_h \sim 48°$ with the mean velocity $v$ of the skyrmion cluster to be $v_x \sim 2.5 \text{ m} \cdot \text{s}^{-1}$ and $v_y \sim 2.8 \text{ m} \cdot \text{s}^{-1}$. To suppress the defect-induced random skyrmion motion, a skyrmion cluster is used here to obtain the reliable current-velocity curve[25]. (**Supplementary Note III**). Intriguingly, the longitudinal velocity of skyrmions is against the electron flow. This means that the minority spin carriers dominate, and the polarization rate $P$ is negative (formula in **Fig. 1a**) in FeGe. First-principles calculations (**Methods** and **Supplementary Note IV**) of the spin-revolved band structure and density-of-state (DOS) in the ferromagnetic phase of FeGe (**Fig. 3a**) support this conclusion. More minority spin (spin-down) bands cross the Fermi level than the majority spin (spin-up) counterparts, and the minority DOS has higher weight at the Fermi level.

Such negative polarization furtherly leads to negative non-adiabatic STT $\beta$, which in turn causes the co-alignment of skyrmion velocity and current-flow direction in 1D case (formula in **Fig. 1b** and **Fig. 2c**). The spin torque is originated from the on-site exchange coupling between the itinerant electrons and localized magnetization[28] via the Hund's coupling $\mathcal{H}_{ex} = -J_{ex}\boldsymbol{\sigma} \cdot \mathbf{M}$ where $J_{ex}$ is the coefficient of on-site exchange interaction, $\boldsymbol{\sigma}$ is the spin of itinerant electrons, and $\mathbf{M} = M_s\hat{\mathbf{m}}$ is the local magnetization with the saturated magnetization $M_s$ unit vector $\hat{\mathbf{m}}$. As shown in **Fig. 3b**, in a ferromagnet, spins of itinerant electrons in the spin-up channel, labelled as $\boldsymbol{\sigma}_+$, are parallel to $\mathbf{M}$ while those in the spin-down channel, labelled as $\boldsymbol{\sigma}_-$, are anti-parallel to $\mathbf{M}$. Both types of itinerant electrons coexist in one phase as long as the DOS at the Fermi level in both spin channels are non-zero. Either $\boldsymbol{\sigma}_+$ or $\boldsymbol{\sigma}_-$ itinerant electrons cost an energy of $2\mathcal{E} = 2|J_{ex}|M$ when it flips its spin directions from its ground state so that $\boldsymbol{\sigma}_\pm$ experiences effective coupling coefficient $J_{ex} = \pm \mathcal{E}/M$, respectively, which can be either positive or negative. By revisiting the Zhang-Li's phenomenological model for non-adiabatic STT and solving the coupled continuity equations for $\boldsymbol{\sigma}_+$ and $\boldsymbol{\sigma}_-$, we obtained the unified formulation of the STT for both adiabatic and non-adiabatic contributions with the effective $\beta = \hbar P/2\mathcal{E}\tau_{sf}$ with reduced Planck constant $\hbar$, spin polarization rate $P$, the spin splitting $2\mathcal{E}$, and spin relaxation time $\tau_{sf}$ (**Supplementary Note V**).

The magnitudes of the Gilbert damping $\alpha$ and non-adiabatic STT $\beta$ are therefore obtained by the study of $v$-$j$ relationship at varied current densities in 1D, the interstitial skyrmions, (**Fig. 3c**) and 2D, skyrmions with ferromagnetic background, (**Fig. 3d**) respectively based on the successful imaging of skyrmion trajectory. According to the linear fitting of $v_x$-$j$, the slope for 1D is $\Delta v_x/\Delta j \approx 8.3$ ($10^{-10} m^3/\text{Å} \cdot \text{s}$), which is about one order of magnitude higher than that of 2D, $\Delta v_x/\Delta j \approx 1.0$ ($10^{-10} m^3/\text{Å} \cdot \text{s}$), demonstrating the high mobility of skyrmions in the 1D system.

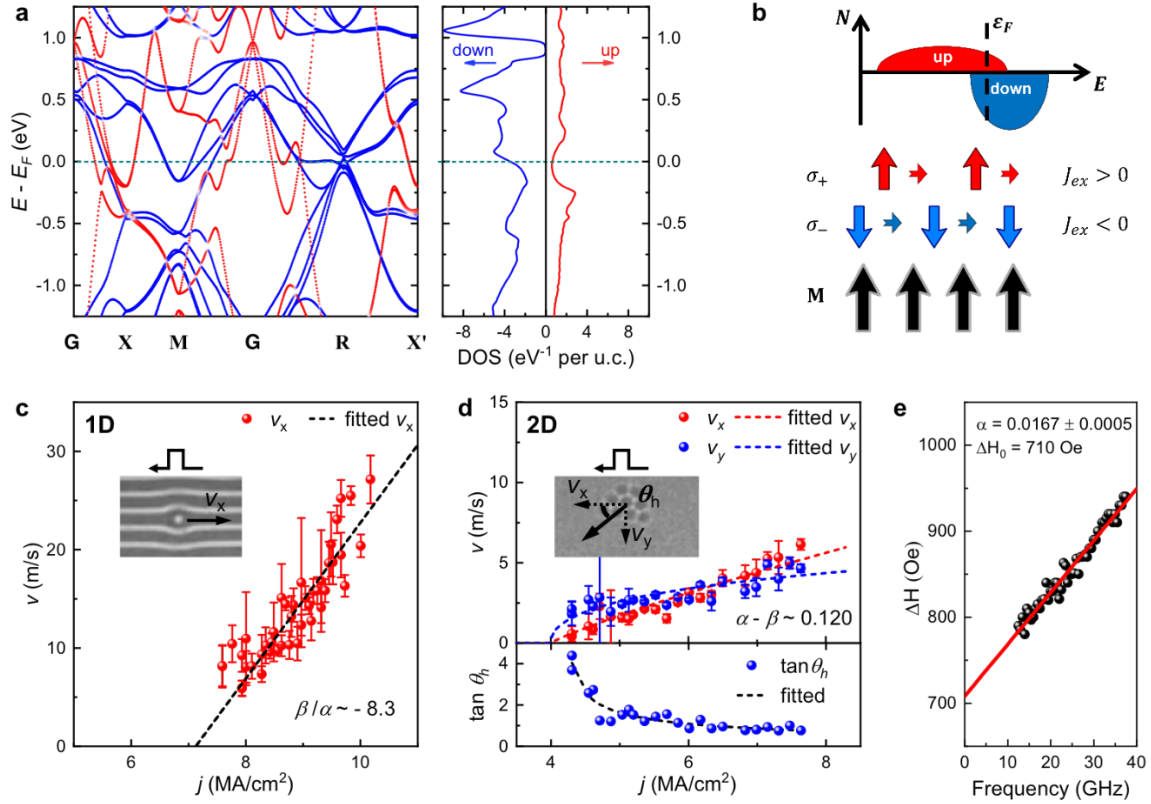

**Fig. 3 | Analysis of the dimension-dependent skyrmion motion. a,** The spin revolved band structure (left) and density-of-state (right) in the ferromagnetic phase of FeGe. Red and blue curve represent spin-up and spin-down channels, respectively. **b,** Schematic of how itinerant electrons of the spin-up ($\sigma_+$) and spin-down ($\sigma_-$) channels in a ferromagnet interact with local spins (**M**) via the positive and negative $J_{ex}$ respectively. **c,** $v_x$-$j$ relationship of the interstitial skyrmion in the helical background. **d,** $v_x(v_y)$-$j$ relationship of skyrmion cluster and the corresponding skyrmion Hall angle $\theta_h$ in the ferromagnetic or conical background. The dashed lines are used to fit the velocity within the theoretical framework of Thiele's equation. **e,** The measured Gilbert damping $\alpha$ of FeGe based on the ferromagnetic resonance. The red solid line is used to fit the data.

In the 2D case, the skyrmion Hall angle of the cluster state shows an inverse relationship $\theta_h \sim 1/j$ with respect to the current density (lower panel of **Fig. 3d**). In particular, large skyrmion Hall angle at low current density signals strong pinning effects in the sample[25]. The relationship between $j$ and the magnitude of $v_x$ is given by $v_x = -\frac{G^2 + \alpha\beta\,D^2}{G^2 + \alpha^2\,D^2} bj \sim -bj$, the solution of the Thiele's equation for skyrmions in the ferromagnetic background (**Supplementary Note I**). Therefore, by using a constant pinning model $\mathbf{F}^{\text{pin}} = -v_{\text{pin}}\hat{v}$ in Eq. (1)[6], best fitting values of Gilbert damping $\alpha$ and non-adiabatic STT $\beta$ to both longitudinal and transverse skyrmion velocities are determined, that is, $\alpha - \beta \sim 0.120$, $\tilde{v}_{pin} = 4.13 \text{ m} \cdot \text{s}^{-1}$, $b = -1.03\,(10^{-10} m^3/\text{Å} \cdot \text{s})$ (Eq. (S7) in **Supplementary Note I**). Accordingly, the spin polarization $P$ is obtained as $-0.683$, a negative value.

In the 1D case, whereas the relationship is given by $v_x = -\frac{\beta}{\alpha}(bj - u_{\text{th}}^{1D})$ where $u_{\text{th}}^{1D} = \frac{1}{\beta D}v_{\text{pin}}$ is the depinning velocity (**Supplementary Note I**). The ratio $\beta/\alpha$ thus governs the skyrmion velocity and its direction, indicating quite different skyrmion dynamics with the 2D case. Through the relationship of $\Delta v_x/\Delta j = -b\,\beta/\alpha$, we straightforwardly get $\beta/\alpha \sim -8.3$, consistent with the inversed sign of $v_x$ in 1D case. Finally, we obtain the values of $\alpha \sim 0.0133$, $\beta \sim -0.107$.

To support these fitting values, a further measurement based on the ferromagnetic resonance gives $\alpha \sim 0.0167$ (**Fig. 3e**, **Methods, Extended Data Fig. 7** and **Fig. 8**), well consistent with the above fitting results. Furthermore, based on the formula of $\alpha$ by Kohno et. al.[29,30] and the newly generic expression of $\beta$, the first-principles calculations suggest $\alpha \sim 0.027$ and $\beta \sim -0.225$ (**Supplementary Note V**). They are both larger than the experimental fitting results, but the calculated $\beta/\alpha = -8.26$ coincides perfectly with the experimental results $\sim -8.3$. The reason is that both $\alpha$ and $\beta$ depend on the spin relaxation time which is environment- and sample-sensitive. In the estimate, we roughly use the electron relaxation time instead. However, the ratio $\beta/\alpha$ is independent on the spin relaxation time $\tau$, since $\tau$ appears in the denominators of the formula for both $\alpha$ and $\beta$. It means that $\beta/\alpha$ is an intrinsic property of the material and dependent on the spin moment $M_\Omega$, the spin polarization $P$, the spin splitting $2\mathcal{E}$ and the PDOS of magnetic atoms at the Fermi level $N_M(\varepsilon_F)$ by (**Supplementary Note V**),

$$\frac{\beta}{\alpha} = \frac{2M_\Omega P}{2\mathcal{E}N_M(\varepsilon_F)} \quad (2)$$

where, $M_\Omega$, $2\mathcal{E}$ and $N(\varepsilon_F)$ are all positive so that the sign of $\beta/\alpha$ is determined by sign of $P$. In order to get large magnitude of $|\beta/\alpha|$, a small product $\mathcal{E}N_M(\varepsilon_F)$ is desired, which is the case for FeGe.

The speed boost of 1D motion in the lane of spin helix turns out to be suitable for various topological defects other than interstitial skyrmion. Thiele's equation neglects detailed structures of topological textures. The same feature would be shared for other defects[23]. Around the geometrical notch, electric current could also excite the disclinations and meron pairs (**Extended Data Fig. 9**). Their current driven motion are shown in **Figs. 4a** and **4b** (**Supplementary Videos 9** and **10**). Under the action of a serial of nanosecond current pulses, the meron pair and disclination are both displaced step by step in a straight line without sideway motion. Under the current density of ~ 9.0 MA·cm$^{-2}$, the velocity of meron pairs and disclinations reaches up to ~ 17.0 m·s$^{-1}$ and ~15.9 m·s$^{-1}$, respectively. The $j$-$v_x$ relations, measured by varying the current density (**Figs. 4c** and **4d**), show a similar window of flow regime as interstitial skyrmion. The $\frac{\Delta v_x}{\Delta j}$ are calculated to be ~ 8.9 and ~ 7.9 for a meron pair and disclinations, respectively, close to the case of interstitial skyrmion. Topological defects have ever been proposed as a new member of data carriers for new concept of spintronic devices[27]. Their high-speed motion under current is unambiguously vital for device applications.

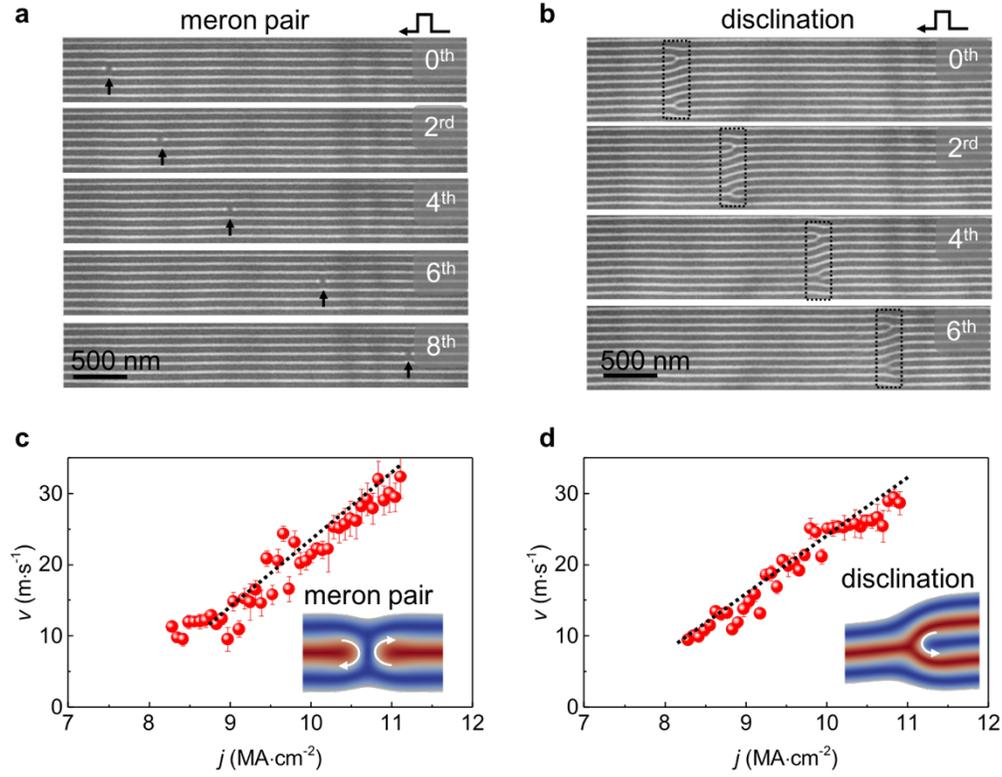

**Fig. 4 | Speed-up effects for other topological defects.** Representative snapshots of the motion of a meron pair in **a** and disclinations in **b** in the spin helix, respectively. The current density is ~ 9.0 MA · cm$^{-2}$ and the duration is 20 ns. The positions of the meron pair and disclinations are indicated with the black arrows and dotted black box, respectively. $j$-$v_x$ relationship of the meron pair in **c** and disclinations in **d**, respectively. The black dotted lines are used to fit the $v_x$ in the flowing regime of the topological defects motion. The insets show the schematic magnetic structures of these topological defects.

In conclusion, we have experimentally explored the 1D skyrmion motion with a vanishing skyrmion Hall angle. The clean helical background and the interstitial skyrmion are created through the well-designed microdevice. The negative spin polarization as well as the negative nonadiabatic STT provides the novel behaviors of skyrmion motion. More importantly, owing to the large magnitude of $\beta/\alpha$, the ratio between two dissipative channels, the nonadiabatic STT and the Gilbert damping, the spatial confinement on skyrmion gives rise to order of magnitude elevation of the slope of current-velocity relationship, compared to that in 2D systems. $\beta/\alpha$ is an intrinsic property of the ferromagnets and depends on the spin moment, the spin polarization, the spin splitting, and the PDOS of magnetic atoms at the Fermi level. These striking features provide an efficient alternative way of manipulating magnetic skyrmions in confined 1D channel. As a new quasiparticle, the interstitial skyrmion could also stimulate future discoveries of composite topological spin textures and inspire a new design of skyrmion-based spintronic devices.

## Methods

### Fabrication of FeGe micro-devices

The FeGe micro-devices tailored for in-situ Lorentz TEM observation were fabricated using the FIB-SEM dual-beam system (Helios NanoLab 600i; FEI) equipped with GIS, and Omniprobe 200+ micromanipulator. Two microdevices with different sizes were used in the experiments to study the current-driven motion of skyrmion in the helical and FM/Conical backgrounds, respectively. The details of in-situ electrical TEM setups for FeGe micro-devices are shown in **Extended Data Fig. 1**.

### In-situ Lorentz TEM experiments of skyrmion motion under electrical current

Magnetic imaging was carried out on the Thermo Fisher Talos F200X operated in the Lorentz Fresnel mode at 200 kV. The objective lens was switched off to provide a magnetic field-free condition. The out-of-plane magnetic field $B$ was pre-calibrated using a Hall effect sensor positioned in a TEM holder. A single-tilt liquid-nitrogen specimen holder (Model 616.6 cryotransfer holder, Gatan) was used with the temperature range from 95 to 380 K. The current pulses were provided by a voltage source (AVR-E3-B-PN-AC22, Avtech Electrosystems Ltd.). The present work sets the electrical current with a pulse duration of 20 ns and a frequency of 1 Hz. The current-driven skyrmion motion was all measured at 95 K.

**First-principles calculations of band structures of FeGe**

Electronic structure of FeGe were calculated from first-principles calculations within the framework of density functional theory (DFT) using the projector augmented wave pseudopotential[31,32] as implemented in VASP[33,34]. The generalized gradient approximation of Perdew, Burke, and Ernzerhof[35] was used for the exchange-correlation energy. Hubbard $U$ with $U = 2\ eV$ in Dudarev approach[36] is applied on Fe(3d) electrons. An energy cutoff 600 eV for the plane-wave expansion was used. Non-collinear magnetism calculations with spin-orbit coupling included were employed. The Γ-centered $k$-point mesh of $9 \times 9 \times 9$ in the Brillouin zone (BZ) were adopted. After we obtained the eigenstates and eigenvalues, a unitary transformation of Bloch waves was performed to construct the tight-binding Hamiltonian in a Wannier function basis by using the maximally localized Wannier functions (WF) method[37] implemented in the Wannier90 package[38]. The WF-based Hamiltonian has the same eigenvalues as those obtained by first-principles calculations from -1.0 ~ 1.0 eV to the Femi level. Transport properties are obtained by using a $64 \times 64 \times 64$ $k$-mesh based on the WF-based Hamiltonian. The further details can be found in Supplementary Note IV.

**Ferromagnetic resonance measurement**

The temperature-dependent damping parameter $\alpha$ was determined by a home-built broadband ferromagnetic resonance (FMR) based on the design[39] shown in **Extended Data Fig. 7a**. A bulk FeGe single crystal was placed onto a co-planar waveguide, and the transmission microwave power as a function of the DC magnetic field was measured. The whole FMR setup was placed into a 9 T - 2 T -2 T superconducting vector magnet with 50 mm bore (Cryogenic, J4440), and the FMR spectra were measured under the field applied perpendicular to the FeGe (111) plane. A typical FMR spectrum measured at 95 K with the microwave frequency of 20 GHz is shown in **Extended Data Fig. 7b**. The resonance signal around 10 - 12 kOe is the uniform FMR mode, which can be well fitted to the generalized FMR line shape[39],

$$\frac{dP}{dH} = A\frac{2\Delta H(H - H_{RES})}{[(H - H_{RES})^2 + (\Delta H)^2]^2} + S\frac{(H - H_{RES})^2 - (\Delta H)^2}{[(H - H_{RES})^2 + (\Delta H)^2]^2} \quad (3)$$

where $H_{RES}$ is the resonance field, $\Delta H$ is the line width, $A$ and $S$ are the amplitude of the anti-Lorentz-type and Lorentz-type line shape. On the other hand, the signal around 6-8 kOe is the edge mode, which is an antiresonance mode due to the coupling between electromagnetic wave and dynamic magnetization at the FeGe surface[40]. Based on the theoretical results[40], the edge mode has smaller amplitude and lower resonance field than that of uniform FMR mode. The FMR spectra have been measured up to 90 kOe, and no significant resonance was witnessed when the applied field is higher than $H_{RES}$.

The damping parameter at 95 K was determined by fitting the frequency dependent resonance linewidth. The FMR data was measured in the frequency range from 11 to 37 GHz with an interval of 0.5 GHz. The selected spectra together with fitting curves of the FMR mode are shown in **Extended Data Fig. 8**. The frequency-dependent linewidth $\Delta H$ shows a linear dependence with frequency, which can be well fitted by

$$\Delta H = \Delta H_0 + \frac{\alpha \omega}{\gamma} = \Delta H_0 + \frac{\alpha f}{\frac{\gamma}{2\pi}} \quad (4)$$

where $\alpha = 0.0152$ is the Gilbert damping parameter, $\Delta H_0 = 0.612$ kOe is the extrinsic linewidth, $f = \omega/2\pi$ is microwave frequency, $\gamma$ is gyromagnetic ratio and $\gamma/2\pi = 2.8$ GHz/kOe was determined form the dispersion fitting.

**Data availability**

The data supporting the findings of this study are available within the article and its Supplementary Information files from the corresponding author upon reasonable request.

## Acknowledgments

H. D. acknowledges the financial support from the National Key R&D Program of China, Grant No. 2022YFA1403603; the Strategic Priority Research Program of Chinese Academy of Sciences, Grant No. XDB33030100; and the Equipment Development Project of Chinese Academy of Sciences, Grant No. YJKYYQ20180012; S. S. P. and J. Z. were supported by the Office of Basic Energy Sciences, Division of Materials Sciences and Engineering, U.S. Department of Energy, under Award No. DE-SC0020221; J. Z. was also supported by Alexander von Humboldt Foundation; A. R. was supported by the Deutsche Forschungsgemeinschaft (DFG) through CRC1238 (Project No. 277146847, project C04); D. S. acknowledges the financial support from the Chinese National Natural Science Foundation (52173215) and the Natural Science Foundation of Anhui Province for Excellent Young Scientist (2108085Y03).

## Author contributions

H.D. supervised the project. J. Z. conceived the theories. H.D. and D.S. conducted the TEM experiments and data analysis. W-S.W. and J. J synthesized FeGe crystals and prepared the TEM samples. J-X.Y. and J.Z. conducted the first-principles calculations. J.Z. and W.W. conducted the micromagnetic simulations. J-X.Y., S.S.P., J.Z. and A.R. conducted phenomenological analysis. P. Z. and L. F. conducted the FMR experiments. H.D., D.S., W.W., J-X.Y., G.Y. and J.Z. prepared the manuscript. All of the authors discussed the results and contributed to the manuscript.

## Competing interests

The authors declare no competing interests.

## Additional information

Supplementary information is available for this paper.

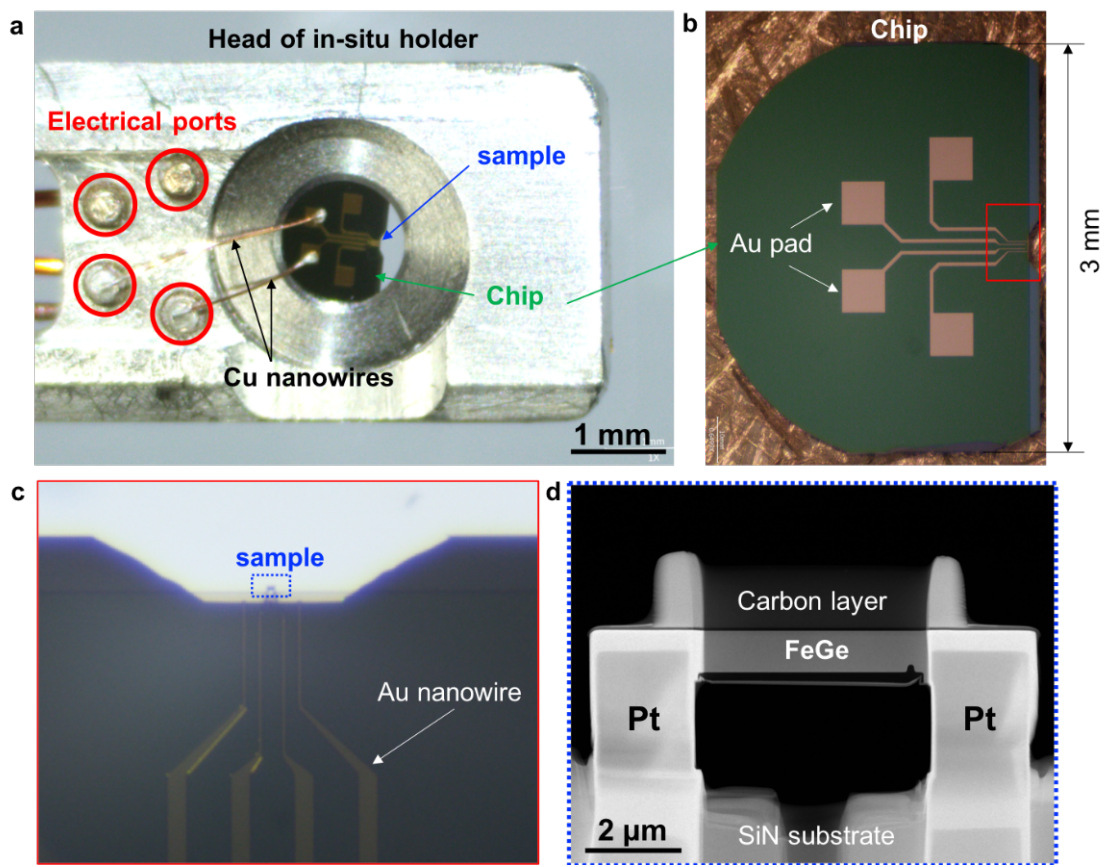

**Extended Data Fig. 1 | In-situ electrical Lorentz TEM experimental setup of FeGe micro-device. a,** The optical image of the head of a Gatan holder, which was designed for in-situ electrical and cooling experiments. The four electrical ports (red circles) were designed by Gatan to be connected to the external voltage source. To build the circuit between the ports and the sample, a customized electrical chip with four Au electrodes was self-designed in **b**. The Au pads were connected to the ports through the Cu wires, which were manually fixed by the silver colloid (only two ports were used in our experiments). The electrical TEM FeGe micro-device was fabricated using focus ion beam (FIB), as such a chip makes the fabrication process compatible with the conventional FIB lift-out method. The enlarge optical image of the FeGe micro-device (red rectangle box in **b**) is shown **c. d,** Low magnification high angle annular dark-field (HAADF) image of the FeGe microdevice (blue rectangle box in **c**) for *in-situ* Lorentz TEM experiments. The FeGe

nanostripe is covered with amorphous carbon on the top and down layers. The left and right edges are connected with two Pt electrodes.

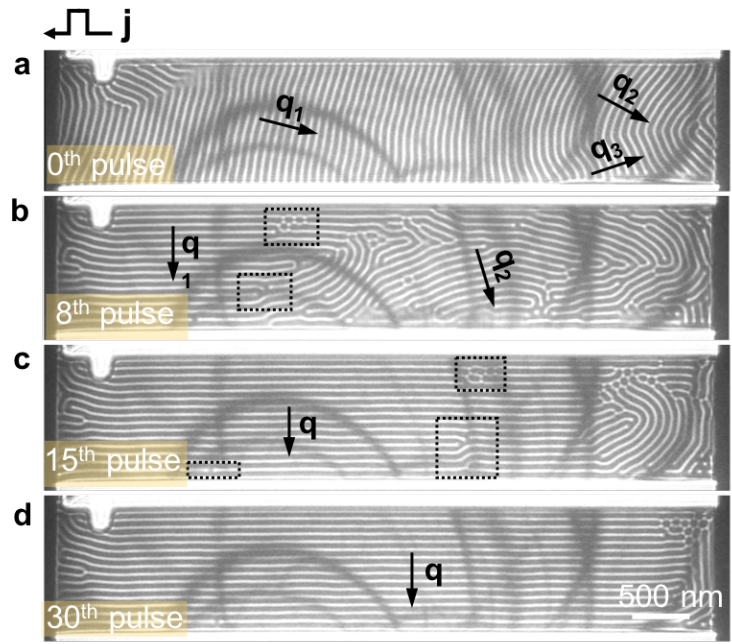

**Extended Data Fig. 2 | Creation of well-defined spin helix. a-d,** Representative snapshots of the generation process of the helical background under the electrical current pulse. The current density is ~ 10.7 $MA \cdot cm^{-2}$. The wavevectors **q** of spin helix are indicated with black arrows. The typical topological defects are marked with black dotted boxes, such as the disclinations, embedded skyrmion clusters and meron pair.

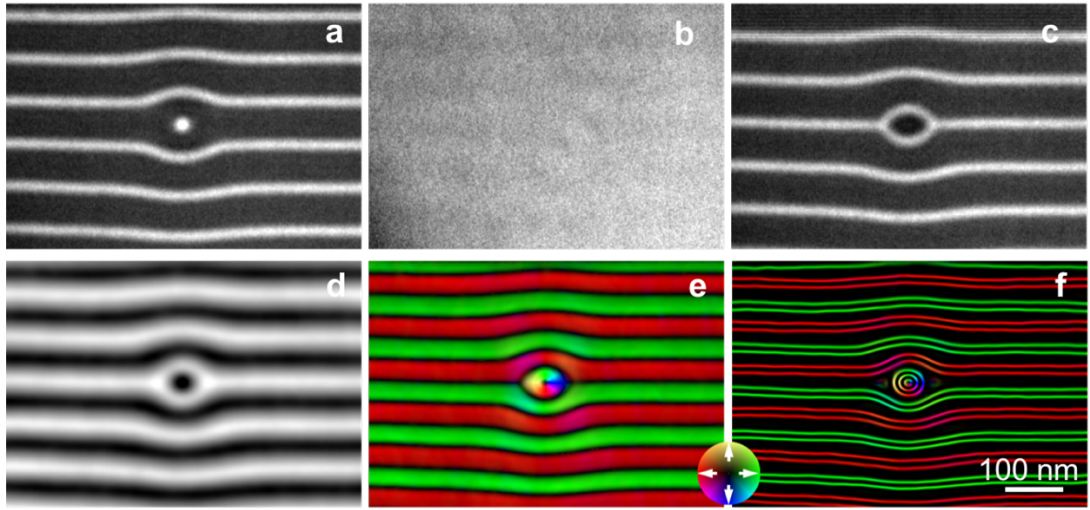

**Extended Data Fig. 3 | Transport of intensity equation (TIE) analysis of the spin textures of the interstitial skyrmion in the helical background**. **a-c,** The defocused LTEM images of the interstitial skyrmion with the defocus value of -400 μm, 0 and 400 μm, respectively. The elliptical interstitial skyrmion and the surrounded distorted helical stripes are clearly observed. **d,** Magnetic phase images extracted by TIE based on **a-c**. **e,** Magnetic induction of the interstitial skyrmion calculated from **d**. **f,** Magnetic contour of the interstitial skyrmion calculated from **d** with the contour spacing of $2\pi/5$.

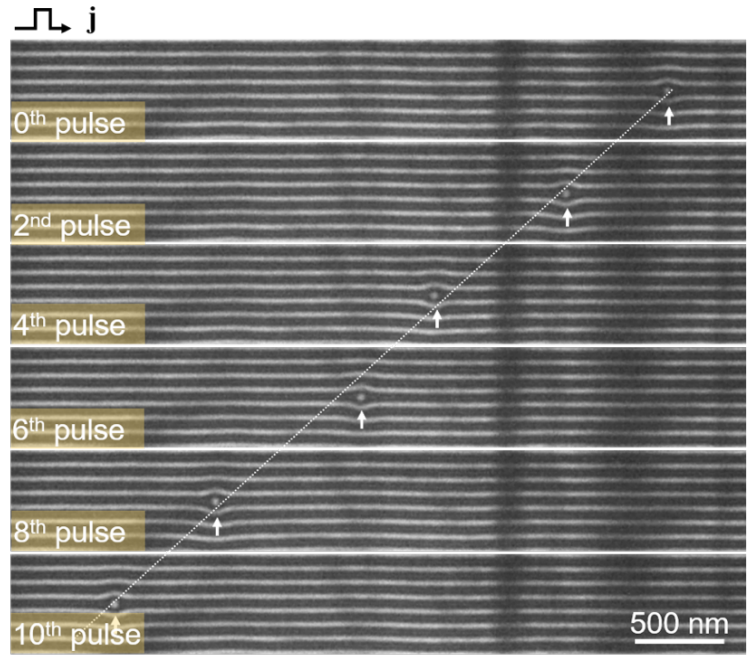

**Extended Data Fig. 4 | Reversal motion of the interstitial skyrmion in the helical background at the current density of ~ 8.6 MA · cm$^{-2}$.** The representative snapshots of the FeGe nanostripe are displayed after applying successive current pulses. The positions of the interstitial skyrmion are indicated with the white arrows. The trajectories of interstitial skyrmion is guided with the white dotted line.

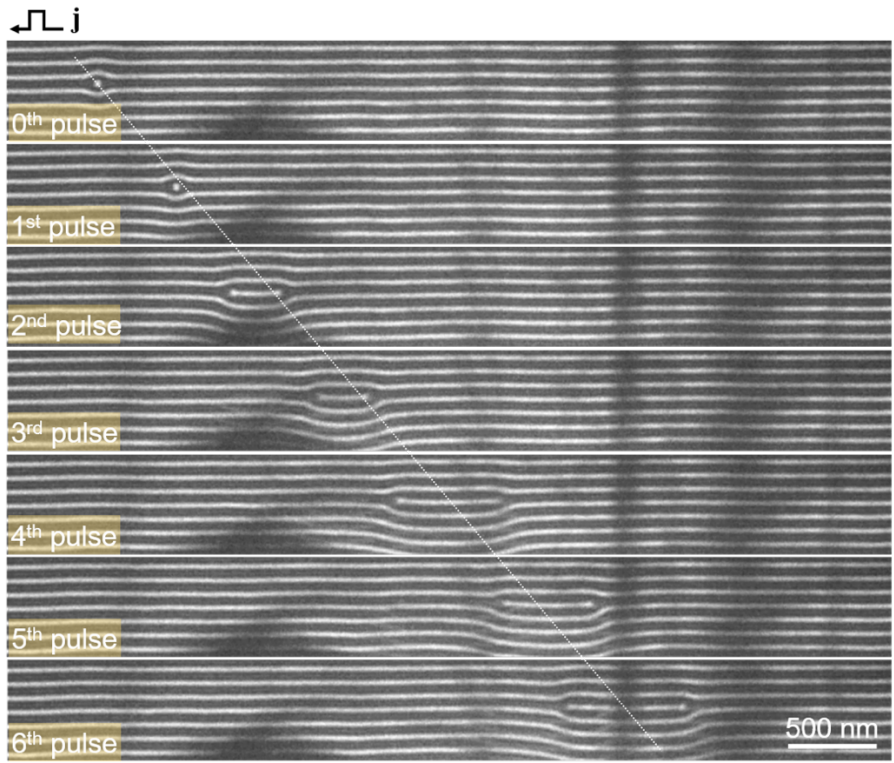

**Extended Data Fig. 5 | Elongation of the interstitial skyrmion at a higher current density of ~ 10.4 MA · cm$^{-2}$.** The interstitial skyrmion is first elongated and then driven forwards under the electrical current. The trajectories of interstitial skyrmion is guided with the white dotted line.

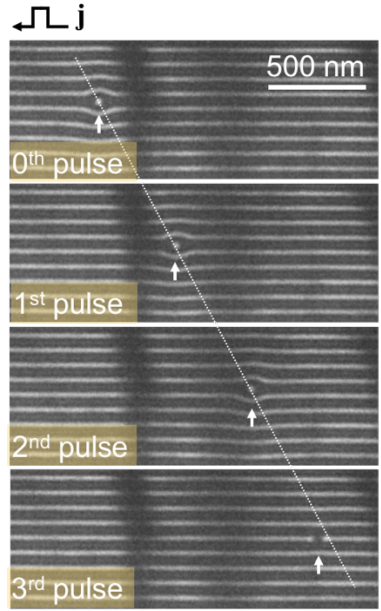

**Extended Data Fig. 6 | The topological structure transition between the interstitial skyrmion and a meron pair at the current density of ~ 10 MA · cm$^{-2}$ in the helical background.** The representative snapshots of the FeGe nanostripe are displayed after applying successive current pulses. The positions of the interstitial skyrmion and a meron pair are indicated with the white arrows. The trajectories of interstitial skyrmion is guided with the white dotted line. The interstitial skyrmion is jumped to the lower helical stripe owing to the skyrmion Hall effect. The transition into a meron pair is owing to the metastability of the interstitial skyrmion in the helical background.

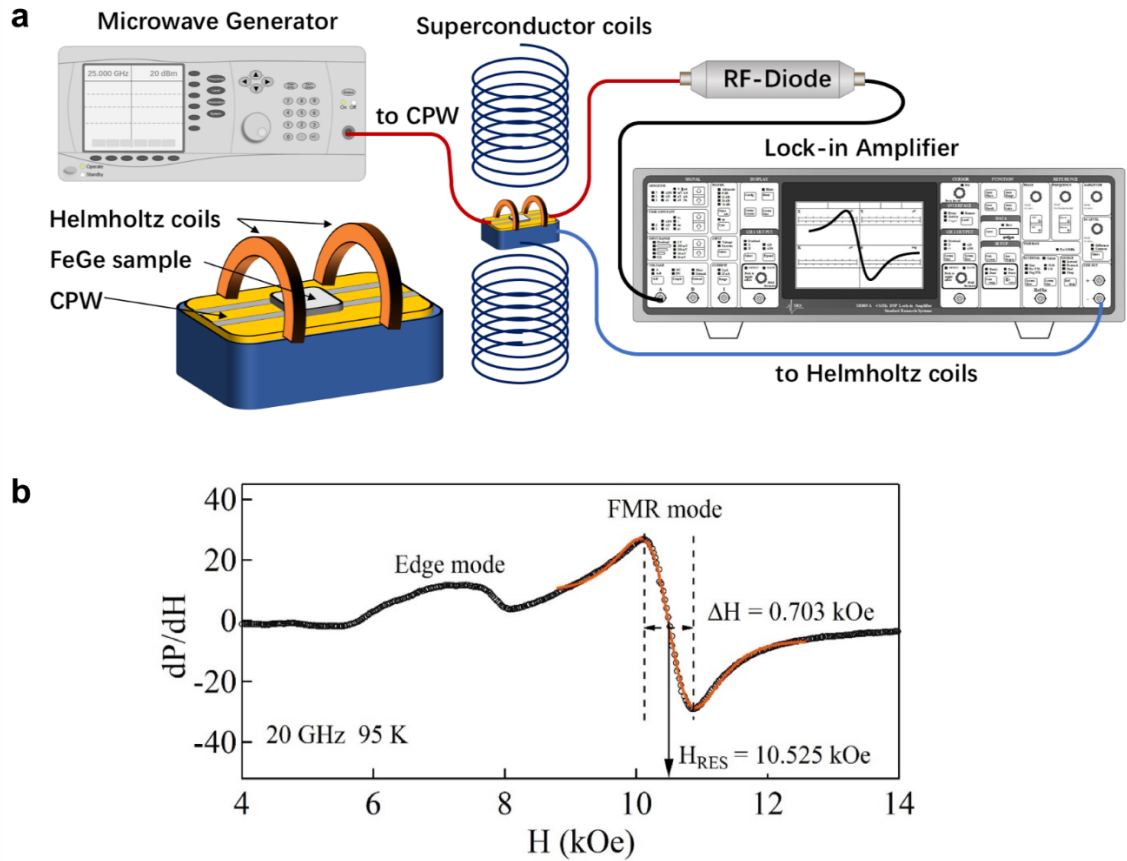

**Extended Data Fig. 7 | The FMR measurement setup. a,** The microwave current which comes from microwave generator was injected into the CPW, and drove the magnetic moment of FeGe processing around the applied magnetic field which generated by a superconducting vector magnet. The transmission microwave was rectified as a DC voltage by the RF-diode, which was detected by Lock-in Amplifier. The Helmholtz coils were used for the modulation of magnetic field, the signal to noise ration therefore can be significantly enhanced. **b,** A typical FMR spectrum of FeGe single crystal at 20 GHz and 95 K with the applied magnetic field applied perpendicular to the FeGe (111) plane. The red curve is used to fit the uniform FMR mode.

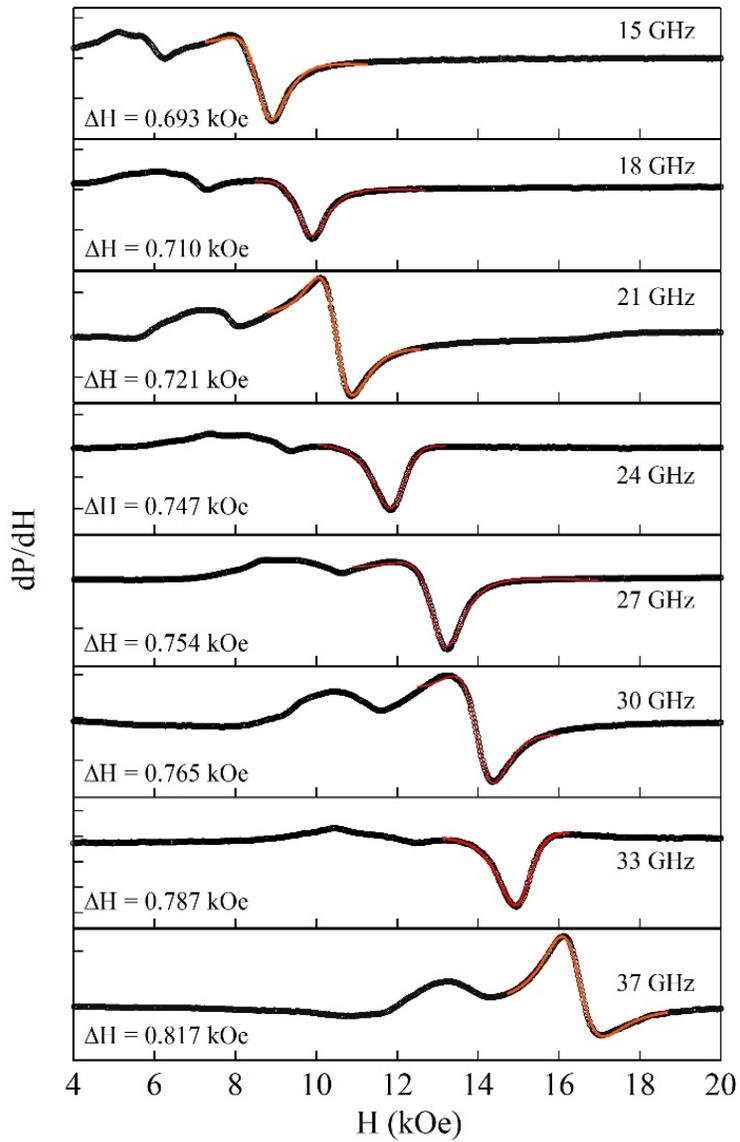

**Extended Data Fig. 8 |** FMR spectra measured at different frequency. The red curves are used to fit the uniform FMR mode.

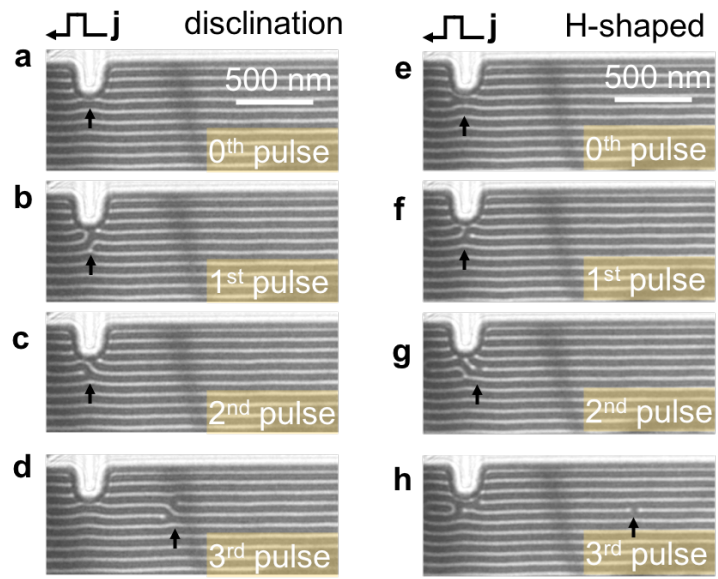

**Extended Data Fig. 9 | Creation of different types of topological defects by geometrical notch in the spin helix at a proper current density of ~ 11 $\text{MA} \cdot \text{cm}^{-2}$. a-d,** Disclination. **e-h,** A meron pair.